\documentclass[aps,prl,superscriptaddress,reprint,floatfix]{revtex4-1}
\usepackage{graphicx,amssymb,amsmath,bm,dcolumn}
\usepackage[caption=false]{subfig}
\usepackage{natbib}
\usepackage{threeparttable}
\usepackage[bottom]{footmisc}
\usepackage{comment}
\usepackage{xfrac}
\usepackage[bordercolor=white,backgroundcolor=gray!30,linecolor=black,colorinlistoftodos]{todonotes}

\usepackage{color}

\begin{document}

\renewcommand{\bibsection}{\section*{Bibliography}}

\title{Ultra-high Q Acoustic Resonance in Superfluid $^4$He}
\author{L. A. De Lorenzo}
\author{K. C. Schwab}
\affiliation{Applied Physics, California Institute of Technology, Pasadena, CA 91125 USA}
\email{schwab@caltech.edu}
\date{\today}

\begin{abstract}
We report the measurement of the  acoustic quality factor of a gram-scale, kilo-hertz frequency superfluid resonator, detected through the parametric coupling to a superconducting niobium microwave cavity. For temperature between 400mK and 50mK, we observe a $T^{-4}$ temperature dependence of the quality factor, consistent with a 3-phonon dissipation mechanism. We observe Q factors up to $1.4\cdot10^8$, consistent with the dissipation due to dilute $^3$He impurities, and expect that significant further improvements are possible.  These experiments are relevant to exploring quantum behavior and decoherence of massive macroscopic objects, the laboratory detection of continuous wave gravitational waves from pulsars, and the probing of possible limits to physical length scales.
\end{abstract}

\maketitle
\section{Introduction}
Recently there has been increasing interest in superfluid optomechanical systems\cite{DeLorenzo2014, Harris2015, Kashkanova2016}.  Superfluid $^4$He offers  extremely low losses in both the mechanical and electromagnetic domains (spanning microwave to optical frequencies). Furthermore superfluid $^4$He is a quantum condensate which can be both cooled deeply below its transition temperature $T_{\lambda}$, $T/T_{\lambda}\approx10^{-2}$ and isotopically purified to extreme purity, realizing very small normal-state fractions of $\rho_n/\rho_0<10^{-8}$, where $\rho_n$ is the density of the normal fluid, and $\rho_0$ is the total density of the fluid.  Together with an ultra-sensitive detection scheme, this system may allow the uniquely quantum behavior which has been observed recently in micromechanical oscillators\cite{oconnell2010, Wollman2015, Riedinger2016} to be revealed in a far larger, gram scale systems.  Furthermore, this system may find utility in diverse applications such as the search for fundamental limits to physical length scales\cite{pikovski2012}, the detection of continuous  gravitational waves\cite{Singh2016}, the understanding of quantum decoherence of massive systems and the emergence of the classical world from the underlying quantum nature\cite{penrose2000wavefunction,ghirardi1986unified,ghirardi1990markov,percival1994primary,fivel1997dynamical,diosi1989models}. 

A key parameter in this systems is the loss rate of the mechanical element.  Here we study the dissipation rate of acoustic modes in $^4$He coupled to a very low loss electromagnetic resonator, and report loss rates which are an order of magnitude smaller than our initial measurements\cite{DeLorenzo2014}, and a factor of more than $10^3$ times less than recent observations in other superfluid optomechanical experiments\cite{Harris2015, Kashkanova2016}.

\begin{figure}[b]
\begin{centering}
\includegraphics[width=.9\linewidth]{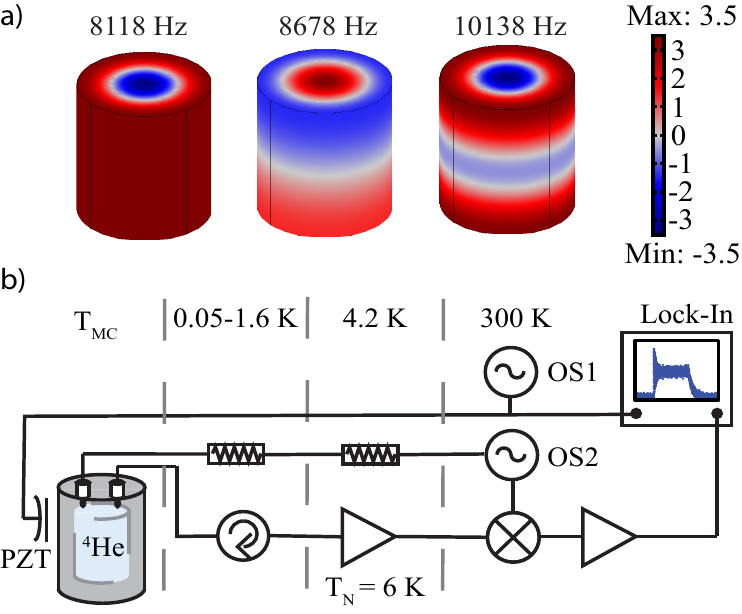}
\end{centering}
\caption{a) finite element simulations of the pressure profiles of the three highest Q acoustic eigenmodes, with mode frequencies shown above each figure. Note that in each of these high Q modes, the radial node is located at the position of the helium fill line. b) Measurement setup: Microwave oscillator OS2 is used to pump the niobium resonator on its red sideband while audio oscillator OS1 drives the piezo at a helium acoustic eigenfrequency.  Parametric coupling between the acoustic and microwave modes results in an upconverted peak at the microwave cavity resonance which is mixed down to an acoustic frequency and measured on a lock-in amplifier.}  
\label{fig1}
\end{figure}

\begin{figure}[t]
\begin{centering}
\includegraphics[width=.85\linewidth]{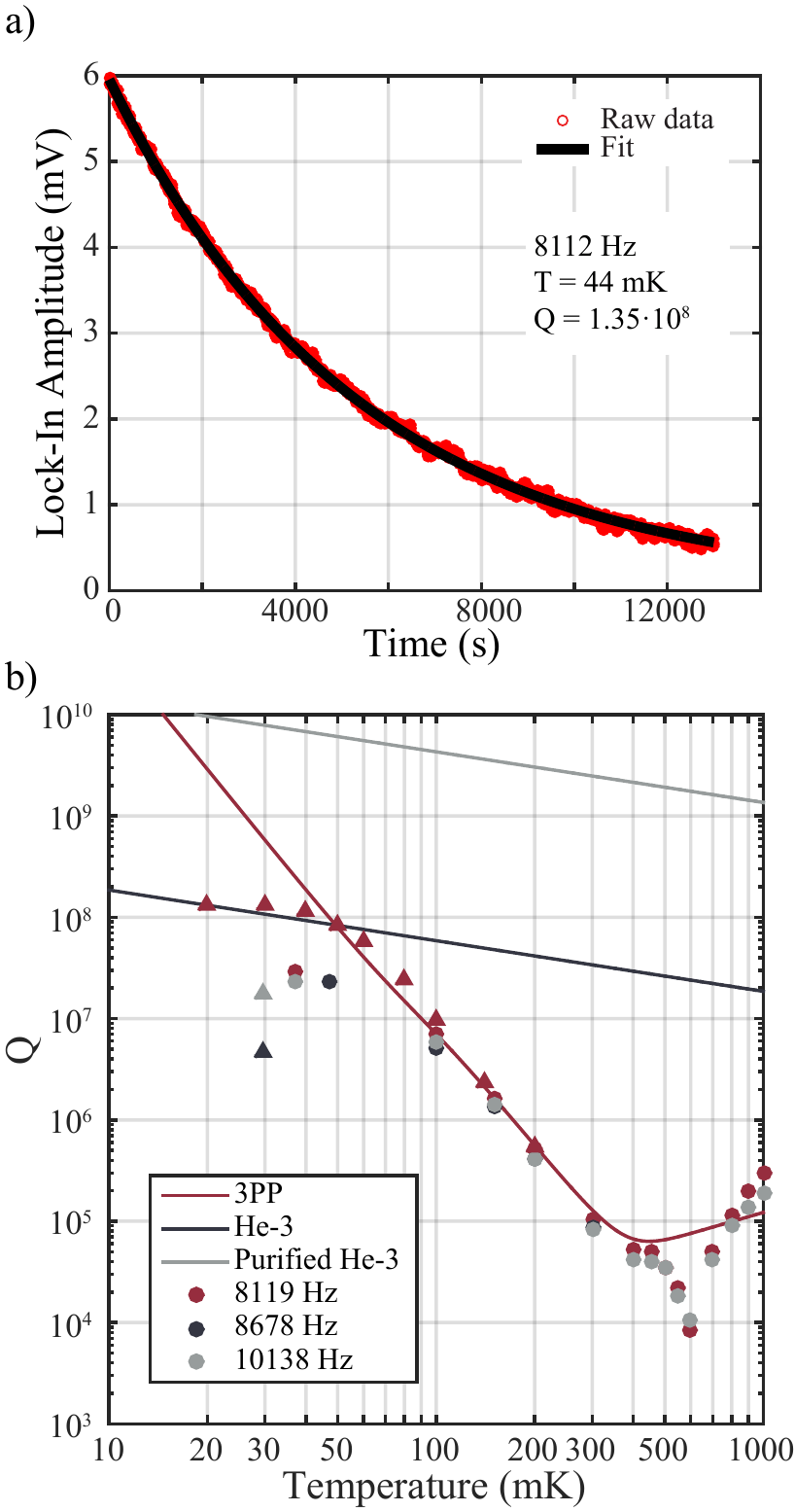}
\end{centering}
\caption{a) is the free decay of the helium acoustic resonance at 8112 Hz, exhibiting a quality factor of $1.35\cdot 10^8$. b) shows quality factor versus fridge temperature for the three highest Q superfluid acoustic modes.  The circles and triangles represent data from our two most recent experimental runs, with improvements specified in the text.  The red line shows the dissipation expected from the 3PP which scales as $T^4$ at low temperatures (Eq. (\ref{eq:1})).  The blue line shows the dissipation expected from $^3$He impurities for concentrations $n_3<10^{-8}$ when the mean free path of the $^3$He atom is less than the container dimensions and the acoustic dissipation is concentration independent, while the gray line show the expected dissipation for $n_3=2 \cdot 10^{-10}$ when the mean free path exceeds the container size by a factor of 100; both cases assume a mode frequency of 8112 Hz.}  
\label{fig2}

\end{figure}

Since the 1950's, acoustic loss in $^4$He has been well studied both experimentally and theoretically\cite{maris1977}.  In the low temperature regime where losses are dominated by the 3-phonon scattering process, the acoustic loss of first sound in helium-4 is given by\cite{abraham1969}
\begin{equation}
\label{eq:1}
\alpha = \frac{\pi^2}{60} \frac{\left(G+1\right)^{2} }{\rho_{4} \hbar^3 c_{4}^{6}} \left(k_{B}T\right)^{4}\omega_{He} \left(\arctan \left(2 \omega_{He} \tau\right) - \arctan\left(\Delta E \tau \right) \right)
\end{equation}
where $G=\left(\rho/c\right)\partial c / \partial \rho=2.84$ is the Gr\"{u}neisen's parameter,\cite{Abraham1970} $\rho=145$ kg/m$^{3}$ and $c_4=238$ m/s are the helium density and speed of sound\cite{Abraham1970}, $\hbar$ is the Planck constant, $k_B$ is the Boltzmann constant, $\omega_{He}$ is the frequency of the acoustic wave, $T$ is the temperature, $\tau=1/(0.9 \cdot 10^7 T^5)$ is the thermal phonon lifetime\cite{jackle1971}, $\Delta E = 3 \gamma \overline{\rho}^2 \omega_{He}$ is the energy discrepancy between the initial and final states in the 3PP, $\overline{\rho} = 3 k_B T/c$ is the average thermal momentum, and $\gamma \approx -10^{48}$ (s/kg$\cdot$m)$^2$ is a constant which characterizes the weak non-linearity of the dispersion relation for low momentum phonons\cite{rugar1984, maris1977}
The loss expected form the 3PP is plotted as in Fig. (\ref{fig2}).  Below $\approx 300$ mK where $\omega \tau >> 1$,  the first $\arctan$ function reaches its maximum value of $\pi/2$ and the loss follows a $T^4$ law, in good agreement with our data.  The second $\arctan$ function is $\approx 0$ down to 100 mK but rises to $\approx -\pi/2$ by 40 mK, increasing the attenuation by a factor of $2$. Eqn. (\ref{eq:1}) predicts a minimum point at $\approx 450$ mK, where $Q$ begins to rise with increasing $T$; our data shows a relatively sharp minimum at 600mK. Furthermore, above $\approx 600$ mK, the roton population is no longer negligible and phonon-roton scattering becomes important, which is not included in Eqn. (\ref{eq:1}).

We form a parametric system coupling low frequency acoustic vibrations in a superfluid filled cavity, approximately 4 cm  in length and 3.6 cm in diameter, to high frequency microwave modes of the surrounding cylindrical niobium resonator\cite{DeLorenzo2014}.  The density modulation produced by the acoustic modes produces a proportional modulation of permitivitty which couples to the microwave modes.  We employ the TE$_{011}$ mode of the microwave cavity, which is typically the highest Q mode in these systems and has a frequency $\omega_C/2\pi = 10.6$ GHz when the resonator is filled with $^4$He.  Input and output coupling to the microwave mode is achieved via two loops of wire recessed into the cavity lid.  The intrinsic loss rate of the TE$_{011}$ mode was measured to be $\kappa_{int}=(2\pi)\cdot 31$ Hz; for the most recent runs of our experiment we have overcoupled the cavity such that $\kappa_{in} = \kappa_{out} = 2 \pi \cdot 230$ Hz . We apply a red detuned microwave pump tone at $\omega_P = \omega_C-\omega_{He}$ while driving the acoustic mode at $\omega_{He}$ with a piezo transducer attached to the niobium cavity, to produce an upconverted sideband at the microwave cavity resonance\cite{Roucheleau2009}.  We detect acoustic modes in the superfluid at frequencies within $1\%$ of their expected values for a right cylindrical acoustic resonator.  Quality factors are determined by recording the free decay of the acoustic oscillations.

In this paper we compare our recent measurements to the initial measurements we reported in Ref.\cite{DeLorenzo2014}.  To improve the mechanical quality factor of the superfluid acoustic modes, we have focused on two known sources of loss: the temperature dependent 3PP described by Eq. (\ref{eq:1}) and the clamping losses from attaching the cell to the dilution refrigerator (DR).  Experimental improvements were identical for both final runs shown in Fig. (\ref{fig2}), with the sole exception being the correction of a thermal short in the fridge which improved the fridge cooling power at each cool stage for the final run (triangles).  

A significant difficulty with very low temperature $^4$He experiments is effectively thermalizing the sample to millikelvin temperatures. Since the speed of sound in $^4$He, $c_4$, is $\approx 1/10$ the speed of sound in most metals, $^4$He has comparatively an extremely high phonon heat capacity at low temperatures.  Furthermore, $^4$He thermalizes only through tranmission of phonons between the fluid and the walls of the container (Nb in this case.) However, due to the severe acoustic impedance mismatch, most of this acoustic energy is reflected resulting in a high boundary resistance,   Kapitza resistance, and is given approximately by: $R_k=15 \hbar^3 \rho_{Nb} c_{Nb}^3/2 \pi^2 k_B^4 T^3 \rho_{4} c_{4} A$, where $\rho_{Nb}$ and $c_{Nb}$ refer to the density and speed of sound in niobium and $A$ is the surface area of contact\cite{pobell2007}.  Since the heat capacity of helium, $C$, depends on temperature as $\propto T^3$ for temperature far below $T_{\lambda}$, we expect the thermal time constant $\tau = R_k C\approx 10$ seconds and to be independent of temperature below $\sim$200 mK (when the $R_k$ is larger than the thermal resistance of the copper wire which suspends the Nb cell from the mixing chamber plate.) However, the very high heat capacity and the convective counter-flows in the fluid lead to a large thermal conductance through the capillaries which fill the cell and connect the helium sample to higher temperature reservoirs.  Together with the high boundary resistance, this leads to substantial difficulty to cool the sample to the base temperature of the dilution refrigerator.  For instance, it would take a heat leak of $\dot{Q}=25nW(0.1nW)$ into the helium resonator to hold the superfluid sample at 40mK (10mK).

In previous runs of the experiment, we observed the helium temperature to be significantly above the fridge base temperature, indicating a failure to sufficiently  thermalize the helium in the fill-line at various refrigerator fridge stages (1K pot, still, cold plate, mixing chamber.)  In the data reported here, we have lowered $R_k$ at three temperature stages by adding silver sinter heat exchangers to the fill line.  Because the acoustic impedance mismatch between helium and a metal is high, large surface areas are required to achieve small values of $R_k$; we have added $6.6$ m$^2$ of surface area to the base plate, and an additional $3.3$ m$^2$ at 100 mK and 975 mK.  By thermally anchoring the helium in the fill line at each stage, heat leaks thru the helium from higher stages of the fridge are reduced.  We have also increased the thermal resistance of the fill line between stages of the DR by using capillaries of smaller diameter ($150-200$ $\mu$m), and longer length ($1$ m).  

However, even with these improvements, we find that at low temperatures, our helium sample has much longer thermalization times than expected.  For instance, in our most recent experimental run, we find $\tau \approx 6$ hours when heating the fridge from 40 to 50 mK, and we note that even at the lowest temperatures, the superfluid $Q$ continued to improve by about $1\%$ per day, suggesting a long term thermal relaxation. 
We have not identified the reason for the long thermal relaxation time.


Our final improvement to reduce the helium sample temperature (triangles only in Fig. (\ref{fig2})) came from increasing the cooling power of our dilution refrigerator by correcting a previously undetected thermal short between 1.6 K and 975 mK.  The result of these changes was an improvement at the mixing chamber to temperatures below 20 mK and an improvement in quality factor of the 8111 Hz mode from $14\cdot10^6$ to $135\cdot 10^6$.  If the acoustic Q is limited by the three phonon process in each case, the helium temperature has decreased from 82mK to 44 mK.  

For mechanical resonators, a significant loss channel is often clamping loss thru the suspension system.  In our most recent experiments, we have decreased the suspension loss by replacing the rigid copper block mounted to cell's midpoint with a 0.13 cm diameter, 6.7 cm length copper wire.  The thermal resistance through this wire at 40mK is expected to be $\approx 10^4 K/W $, a factor of 100X less than the Kapitza resistance between the helium and the niobium walls.    We have also replaced the lid of the cell, moving the helium fill line and microwave couplers to the location of the first radial acoustic node, thereby reducing acoustic loss from acoustic radiation into the fill line.

Finally, we note that acoustic loss is expected to arise from the dilute $^3$He impurity, which behaves as a classical viscous gas in the superfluid\cite{kerscher2001}.  Following this assumption we calculate the acoustic attenuation from $^3$He (at natural concentration of $\approx$ 1 in  $10^6$ helium atoms) to be\cite{DeLorenzo2015}:
\begin{equation}
\label{eq:a_he3_di}
\alpha_{3He} \approx \left( \frac{7}{3} \sqrt{\frac{k_{B} m^{*}_{3}}{\pi}} \frac{1}{\sigma}\right) \left( \frac{1}{ \rho_{4} c_{4}^{3}} \right) \left(\sqrt{T} \omega^{2} \right)
\end{equation}
where $\sigma = \pi d^{2}=6\cdot 10^{-20}m^2$ is the He-3 scattering cross section and $m^{*}_{3} = 2.34m_{3}$ is the effective mass of a $^{3}$He atom at zero concentration\cite{bardeen1966}.  The expected loss from $^3$He is plotted as a blue line in Fig. (\ref{fig2}).  The losses due to $^3$He are expected to be independent of concentration until the mean free path of the $^3$He atom exceeds the container dimensions; in this case, for concentrations above $x=n_3/n_4>10^{-8}$, where $n_3$ and $n_4$ are the number density of $^3$He and $^4$He atoms.  The loss rate from $^3$He can be diminished by decreasing the $^3$He concentration with isotopically purified samples of helium\cite{DeLorenzo2015}.  In our most recent experimental run, the quality factor of the 8.119 kHz mode has become comparable to the $^3$He impurity limit at low temperatures.   In an attempt to improve the Q further, we warmed the system to above 4 K, flushed the cell and refilled it with a sample of helium with only $2\cdot10^{-10}$ $^3$He concentration\cite{mcclintock1978}, but did not note an improvement in quality factor of the 8 kHz mode.  We did not measure the isotopic purity of the helium removed from the cell following the experiment and are unable to confirm that we adequately removed the $^3$He impurity from the setup.

The low temperature, the highest acoustic Q data points were taken with an incident microwave power at the niobium cavity of $0.4$ pW, corresponding to $3\cdot10^4$ pump photons in the cavity.  We are able to apply powers up to $4$ nW or $3\cdot10^8$ pump photons in the cavity while maintaining a superfluid acoustic $Q$ above $10^8$.  Higher powers begin to heat the helium and degrade the mechanical $Q$.  For instance at $40$ (400) nW, the helium temperature extracted from the $Q$ is 61 (122) mK.  This level of heating is much higher than expected given the low dielectric loss tangent of helium $tan\left(\delta\right) < 10^{-10}$\cite{hartung2006} and of our niobium cell $1/Q = 5\cdot 10^{-8}$.  A possible explanation is the normal metal remaining in our microwave circuits: the coupling loops embedded in the Nb lid are BeCu and the SMA caps are Cu; in future runs of this experiment these will be replaced with superconducting materials.

Reaching and acoustic $Q>10^{10}$ will require lowering the helium temperature from our current $44$ mK to $<15$ mK.  To reduce the heat leak from the fill-line to the cell we will install a superfluid leak tight cryogenic valve located on the mixing chamber\cite{bruckner1996}.  With the valve in place, the fill line from the mixing chamber to room temperature can be evacuated to vacuum and heat leaks resulting from helium film flow up the capillaries will be eliminated.  We can also replace the copper wire used for suspension with a high purity (5N) annealed silver wire, which due to higher thermal conductance will allow for an even smaller diameter wire to be used, minimizing further the mechanical contact to the cell.  

Reaching such high quality factors is likely to require further limiting clamping losses.  In our current setup, there are four mechanical connections to the cell: the suspension and cooling wire, the fill line, and two microwave connections.  A simple improvement in suspension loss can be accomplished by utilizing a higher Q material (such as silver) for the suspension wire,\cite{braginsky1985}, and to engineer the vibrational modes of the suspension such that there are no modes at the helium acoustic frequency.  It is possible to eliminate one microwave connection by operating the niobium cavity in reflection or to eliminate both microwave connections by using antenna coupling\cite{ivanov1993}.  The fill line could be removed by welding the cavity lid in place and pre-filling the cell to a pressure of $2.3\cdot10^{7}$ Pa (230 bar) at 77 K or $9\cdot10^{7}$ Pa (900 bar) at 300 K.  Alternatively, the fill line and suspension wire can be combined to a single connection.  

The viscous loss expected from the $^3$He impurity can be lowered by using isotopically purified samples of helium.  To reach $Q>10^{10}$ in our system requires a $^3$He concentration of $x =2\cdot10^{-10}$, (see Fig. (\ref{fig2})) which we have already purchased.\cite{mcclintock1978}  Samples with $^3$He concentrations as low as $x =10^{-12}$ are available\cite{hendry1987continuous}.

Finally we note that prospects for measurements of the thermal motion of the acoustic modes appears well within reach.  Assuming a microwave cavity with $Q=10^9$, negligible heating of the helium from dielectric loss, and an acoustic quality factor limited by the three phonon process (Eqn. (\ref{eq:1})), it will be possible to measure the thermal motion of the mode at $14$ ($8$) mK with a precision of $10\%$, where $Q = 10^{10}$ ($Q = 10^{11}$) with a source of phase noise $-143$ ($-136$) dBc/Hz at an 8 kHz offset from carrier.  This level of phase noise is possible with demonstrated microwave sources\cite{ivanov1998microwave}

In conclusion, we have realized a superfluid $^4$He acoustic resonator with exceedingly high quality factor, $Q=1.4\cdot 10^8$, which for the lowest dissipation mode appears to be limited by the dissipation due to dilute $^3$He impurities.  Only three other materials (silicon, sapphire, and quartz)\cite{bagdasarov1977,mcguigan1978,goryachev2012} have ever shown lower acoustic losses, with loss rates approximately 10x lower than what we demonstrate here.  Given the unquie properties of superfluid $^4$He it appears possible for $^4$He to achieve significantly lower loss rates. Further development of this unique system will find applications which span from the detection of continuous gravitational waves from sources such as nearby pulsars\cite{Singh2016}, quantum decoherence of macroscopic systems\cite{penrose2000wavefunction,ghirardi1986unified,ghirardi1990markov,percival1994primary,fivel1997dynamical,diosi1989models}, to studies which search for limits to physical length-scales at extremely small distances\cite{pikovski2012}.


\begin{acknowledgments}
We acknowledge funding provided by the Institute for Quantum Information and Matter, an NSF Physics Frontiers Center (NSF IQIM-1125565) with support of the Gordon and Betty Moore Foundation (GBMF-1250) NSF DMR-1052647, and DARPA-QUANTUM HR0011-10-1-0066.  L.D. acknowledges support from the NSF GRFP under Grant No. DGE-1144469.
\end{acknowledgments}

\bibliographystyle{apsrev4-1}
\bibliography{bib2016}

\end{document}